\documentclass{andromedaone}       

\usepackage[utf8]{inputenc}
\usepackage{feynmf}
\usepackage{feyn}
\usepackage{csquotes}
\usepackage{slashed}

\usepackage{rotating}
\usepackage{adjustbox}
\usepackage{xcolor}
\usepackage{mathtools}

\usepackage{float}
\usepackage{graphicx}

\usepackage{amsmath,tikz}
\usepackage[makeroom]{cancel}
\usepackage{empheq}

\usetikzlibrary{matrix,calc,decorations.markings}
\usepackage{amssymb}
\usepackage{amsfonts}

\journal{BSM}
\vol{2021}
\jyear{Egypt}
\pages{(CFP) Zewail City of Science and Technology} 

\received{xx April 2021}
\published{xx May 2021}

\def\be{\begin{equation}}
\def\ee{\end{equation}}
\def\bea{\begin{eqnarray}}
\def\eea{\end{eqnarray}}

\newcommand{\nn}{\nonumber}

\newcommand{\GeV}{\; \mathrm{GeV}}
\newcommand{\TeV}{\; \mathrm{TeV}}
\newcommand{\beq}{\begin{equation}}
\newcommand{\eeq}{\end{equation}}

\newcommand{\grav} {\widetilde{G}}

\newcommand\iso[2]{\mbox{${}^{#2}${\rm #1}}}

\def\be#1{\iso{Be}{#1}}

\def\b1#1{\iso{B}{1#1}}

\makeatletter

\def\slash{\@ifnextchar[{\fmsl@sh}{\fmsl@sh[0mu]}}
\def\fmsl@sh[#1]#2{%
  \mathchoice
    {\@fmsl@sh\displaystyle{#1}{#2}}%
    {\@fmsl@sh\textstyle{#1}{#2}}%
    {\@fmsl@sh\scriptstyle{#1}{#2}}%
    {\@fmsl@sh\scriptscriptstyle{#1}{#2}}}
\def\@fmsl@sh#1#2#3{\m@th\ooalign{$\hfil#1\mkern#2/\hfil$\crcr$#1#3$}}
\makeatother

\def\b{\beta}

\def\d{\delta}

\def\g{\gamma}

\def\m{\mu}
\def\n{\nu}

\def\r{\rho}

\def\x{\xi}

\newcommand{\mplanck}{\ensuremath{M_{\text{P}}}}

\def\slash#1{\rlap{\hbox{$\mskip 1 mu /$}}#1}   

\makeatletter 
\def\slash{\@ifnextchar[{\fmsl@sh}{\fmsl@sh[0mu]}} 
\def\fmsl@sh[#1]#2{%
  \mathchoice 
    {\@fmsl@sh\displaystyle{#1}{#2}}%
    {\@fmsl@sh\textstyle{#1}{#2}}%
    {\@fmsl@sh\scriptstyle{#1}{#2}}%
    {\@fmsl@sh\scriptscriptstyle{#1}{#2}}} 
\def\@fmsl@sh#1#2#3{\m@th\ooalign{$\hfil#1\mkern#2/\hfil$\crcr$#1#3$}} 
\makeatother 

\definecolor{darkgreen}{rgb}{0,.5,0}

\newcommand{\sg}{{\tilde g}}
\newcommand{\sq}{{\tilde q}}
\newcommand{\Gr}{  {\grav}}

\allowdisplaybreaks
\usepackage{bbold}

\begin{document}

\title{Gravitino thermal production}

\author{Helmut~Eberl,\auno{1} Ioannis~D.~Gialamas,\auno{2} and Vassilis~C.~Spanos \auno{2} }
\address{$^1$ Institut f\"ur Hochenergiephysik der \"Osterreichischen Akademie der Wissenschaften, \it A--1050 Vienna, Austria}
\address{$^2$ National and Kapodistrian University of Athens, Department of Physics, Section of Nuclear {\rm \&} Particle Physics,  GR--157 84 Athens, Greece}

\begin{abstract}
In this talk{\footnote{Invited talk given at BSM--2021 by V.C. Spanos based on~\cite{Eberl:2020fml}.}} we present a new calculation of the gravitino   production  rate,  using  its full one-loop corrected thermal   self-energy, beyond the hard thermal loop approximation. Gravitino production $2 \to 2$ processes, that are  not related to  its  self-energy   have been taken properly   into account.
Our result, compared to the latest  estimation, differs  by  almost 10\%. 
In addition, we present a handy parametrization  of our finding, that can be used 
to calculate the gravitino thermal   abundance, as a function of the reheating temperature.
\end{abstract}

\maketitle


\section{  INTRODUCTION}
Gravitino,  the  superpartner of graviton,  naturally belongs to the 
spectrum of models that extend the Standard Model (SM)
in the context of  supergravity (SUGRA). This particle  can play the 
role of the dark matter (DM) candidate particle and since it 
 interacts  purely 
 gravitationally with the rest of the spectrum,  naturally escapes 
 direct or indirect detection, as the current   experimental and observational data 
 on DM searches  suggest.
  Therefore, the precise knowledge of its cosmological  abundance is  essential    to apply cosmological constraints on these models.  
Gravitinos can  be produced: (i) nonthermally,  
 from the  inflaton decays~\cite{Kallosh:1999jj,Giudice:1999am,Nilles:2001ry,Kawasaki:2006gs,Endo:2006qk,Ellis:2015jpg,Dudas:2017rpa,Kaneta:2019zgw,Garcia:2020wiy}, (ii)  much  later around  the big bang nucleosynthesis  epoch, through  
 the decays of   unstable particles~\cite{Kawasaki:2008qe,Kawasaki:2017bqm,Cyburt:2006uv,Cyburt:2012kp}  and (iii)   thermally, using a freeze-in production 
 mechanism, as the Universe cools  down from the  
  reheating temperature ($T_\mathrm{reh}$) until now~\cite{Weinberg:1982zq,Ellis:1984eq,Khlopov:1984pf,Moroi:1993mb,Kawasaki:1994af,Moroi:1995fs,Ellis:1995mr,Bolz:1998ek,Bolz:2000fu,Bolz:2000xi,Steffen:2006hw,Pradler:2006qh,Pradler:2006hh,Rychkov:2007uq,Pradler:2007ne,Ellis:2015jpg,Eberl:2020fml}. 
Assuming    gauge mediated supersymmetry breaking,
 a different production mechanism (freeze-out)  is possible~\cite{Giudice:1998bp,Choi:1999xm,Asaka:2000zh,Jedamzik:2005ir}.
Recently, an alternative scenario, where the so-called  
``catastrophic" non-thermal production of slow gravitinos 
has attracted  attention~\cite{Kolb:2021xfn,Kolb:2021nob,Dudas:2021njv, Terada:2021rtp,Cribiori:2021gbf,Castellano:2021yye}.

The first attempt to calculate the   gravitino thermal abundance, is already 
almost forty years old. 
Since the gravitinos are   thermally produced at very high temperatures, the effective 
theory  of light gravitinos, 
the so-called nonderivative approach, involving only the spin  1/2 goldstino  components, was initially used.  
In this context,  as some of the production amplitudes exhibit infrared (IR) divergences,  they were regularized 
by introducing either a finite thermal gluon mass or an angular cutoff.
Thus,  in~\cite{Ellis:1984eq} the basic $2 \to 2$ gravitino production  processes
    had been   tabulated and  calculated for the first time, see Table~\ref{table:1}. 
Afterwards, this calculation was further improved in~\cite{Moroi:1993mb, Kawasaki:1994af}.  Since the  Braaten, Pisarski, Yuan (BPY) method~\cite{Braaten:1989mz,Braaten:1991dd}
 succeeded in calculating the axion thermal abundance, in~\cite{Ellis:1995mr} it was also applied 
 to the gravitino,  motivated by the fact that  the gravitino like axion, 
 interacts extremely weakly with the rest of the spectrum. 
Although in~\cite{Bolz:1998ek} the previous  IR regularization technique  was used, in~\cite{Bolz:2000fu,Pradler:2006qh}  the BPY method was employed, 
calculating also  the contribution of the spin 3/2 pure gravitino components. In~\cite{Rychkov:2007uq} a different   calculation method was used. 
In this paper  it was  argued that   the basic requirement to apply
 the BPY prescription, i.e.  $g \ll 1$, where  $g$ is the gauge coupling constant,
can not be fulfilled   in the whole temperature range of the calculation, 
especially  if  $g$ is the strong coupling constant $g_3$. 
 Therefore, the authors  calculated   the   one-loop  thermal gravitino self-energy    numerically      beyond the hard thermal loop  approximation. 
 It is important to note that that calculation   incorporates in addition  the  $1 \to 2$ processes.
It was also noticed that the so-called  subtracted part,  i.e.  pieces  of the   $2 \to 2$ squared amplitudes
   for which the self-energy  may  not  account for,  {\em are IR finite}. 
 The main numerical result in~\cite{Rychkov:2007uq} on  the gravitino production rate
   differs significantly, almost by a factor of 2, with respect to the previous works~\cite{Pradler:2006qh,Pradler:2007ne}, see Fig.~\ref{fig:gamma}.
 Unfortunately, in~\cite{Rychkov:2007uq} basic equations 
 appear to be inconsistent,  in particular   in Sec. IV A
 the equations on the self-energy contribution for  the gravitino production.
Moreover, the  numerical estimation of this self-energy was computed only  inside the light cone   due to  the limited computation resources at  that time.  
In addition, two  out of the four nonzero  subtracted parts in the corresponding  Table I in~\cite{Rychkov:2007uq},   turn out to be zero.

Motivated by these, in this talk  we present a new calculation of
the thermally corrected gravitino self-energy.
Eventually, for our final result  for the gravitino production  rate
we present an updated  handy    parametrization of this. 
Our  result differs  from that shown in~\cite{Rychkov:2007uq} approximately by  10\%.
Finally, we present  the gravitino thermal abundance and discuss possible phenomenological consequences.  
\section{The calculation of the thermal rate}
\begin{figure}[t]
\centering
\includegraphics[width=0.9\textwidth]{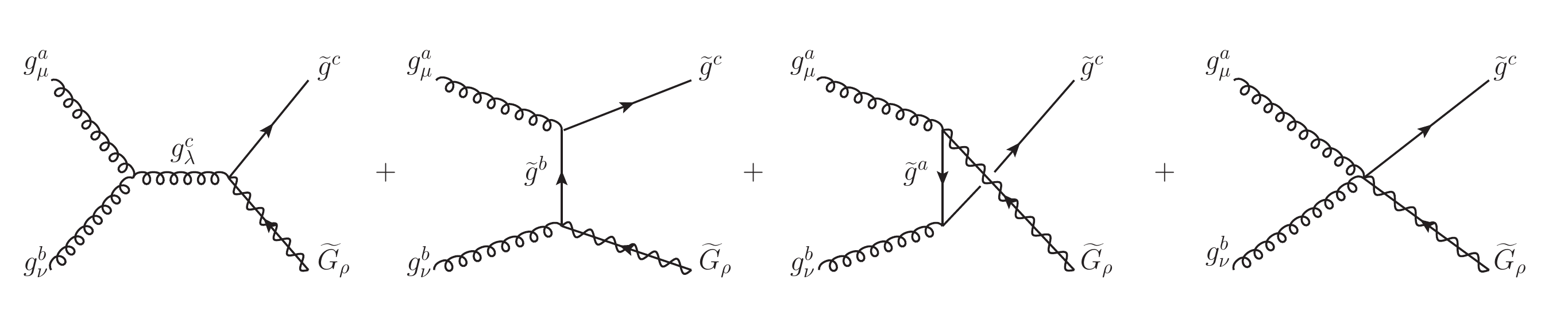}\\
\includegraphics[width=0.5\textwidth]{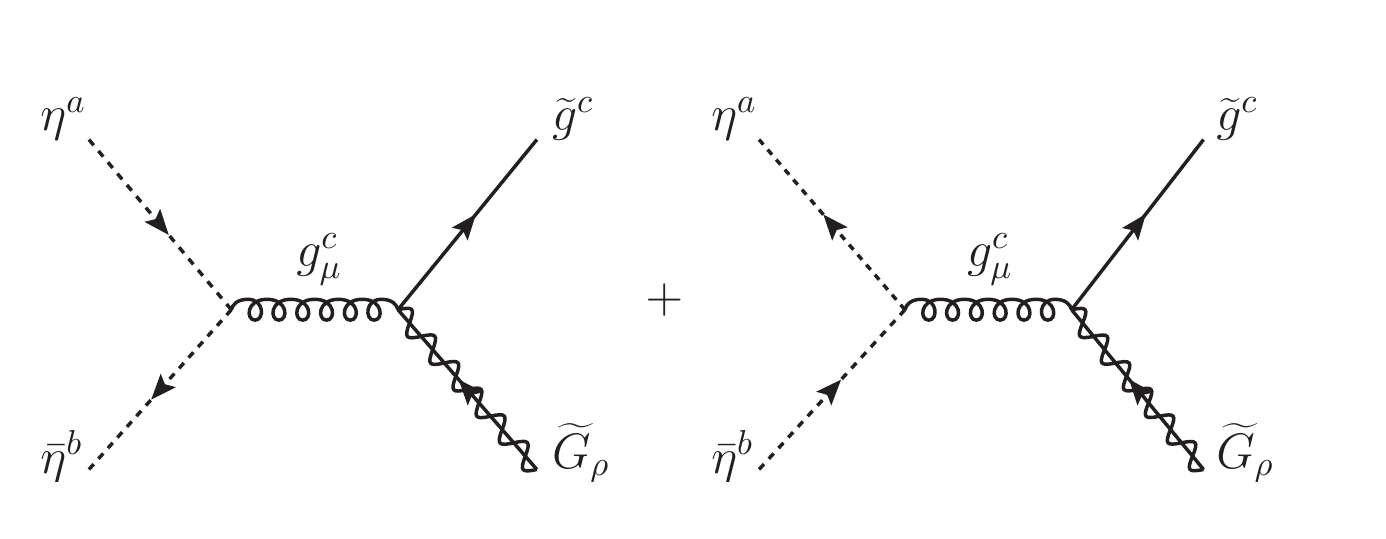}
\caption{The relevant diagrams for the process $g g \to \sg \Gr$. In the top row are displayed in turn the $s$, $t$, $u$ channels and the quartic  vertex $x$. In the bottom row the Faddeev–Popov ghosts diagrams are shown. Although here we present only the  $SU(3)_c$ case, the same diagrams appear also for the $SU(2)_L$ gauge group, but not for the $U(1)_Y$.  }
\label{fig:process_A}
\end{figure}
As the gravitino is the superpartner of the graviton, 
its interactions are suppressed by the inverse of the reduced Planck mass  $\mplanck=(8\pi\, G)^{-1/2}$. 
Hence, the dominant contributions to its production, in leading order of the gauge group couplings, are  
 processes of the form $a\,b \rightarrow c\, \widetilde{G}$,  where $\widetilde{G}$ stands for gravitino and  $a$, $b$,  $c$ can be 
three superpartners or one superpartner and two SM particles.
The possible processes and the corresponding  squared amplitudes in
$SU(3)_c$ are given in Table \ref{table:1}, where for their denotation by the letters A-J we follow the ``historical" notation of~\cite{Ellis:1984eq}.
\begin{table}[h]
\tbl{ Squared matrix elements for  gravitino  production in $SU(3)_c$ in terms of $ g_3^2 \, Y_3/\mplanck ^2$  assuming massless particles, with  $Y_3 =  1 +m^2_{\sg}/(3 m^2_{\tiny 3/2} )$. The Casimir operators are $C_3 =   24$ and $C'_3 =   48 $. \label{table:1} }
{\begin{tabular}{cccc}
\hline \hline \\[-3mm]
$X $ & Process &  $ |{\cal M}_{X,\rm full}|^2  $ &  $ | {\cal M}_{X,\rm sub}|^2 $  \\[1.2mm]
\hline \\[-3mm]
 A & $g g \to \sg \Gr$ & $4 C_3 ( s + 2 t + 2 t^2/s) $  & $- 2 s C_3 $\\
 B & $   g \sg \to g \Gr$   & $ - 4 C_3 (t + 2 s + 2 s^2/t)$  & $  2 t C_3$  \\
 C & $ \sq g \to q \Gr  $ &  $ 2 s C'_3 $ &  $0 $\\
 D & $ g q \to \sq \Gr $ & $ - 2 t C'_3 $ & $ 0$ \\
 E &  $   \sq q \to g \Gr $  & $- 2 t C'_3 $ & $ 0 $ \\
 F & $\sg \sg \to \sg \Gr $& $8  C_3 (s^2 + t^2 + u^2)^2/(s t u)  $ & $ 0$ \\
 G & $ q \sg \to q \Gr $& $- 4 C'_3 (s + s^2/t) $ &  $0$ \\
 H  & $  \sq \sg \to \sq \Gr $ & $-2 C'_3 (t + 2 s + 2 s^2/t)  $   &  $0 $\\
 I & $q \sq \to \sg \Gr $ & $- 4 C'_3 (t + t^2/s)  $& $ 0$ \\
 J &  $ \sq \sq \to \sg \Gr $ & 2 $ C'_3 ( s + 2 t + 2 t^2/s) $ & $ 0$ \\[1mm]
 \hline \hline 
\end{tabular}}
\end{table}
In  $SU(3)_c$, the particles $a$, $b$, and $c$ could be gluons $g$, gluinos $\sg$, quarks $q$, or/and squarks $\sq$. 
Analogous processes happen in $SU(2)_L$ or $U(1)_Y$, 
where the gluino  mass $m_\sg \equiv M_3$ becomes  $M_2$ or $M_1$, 
respectively.
In the factor $Y_N \equiv   1 + { m_{\lambda_N}^2 /  (3 m^2_{\tiny 3/2})}$,  
where $m_{\lambda_N} = \{ M_1, M_2, M_3 \}$ and $ m_{\tiny 3/2}$ is the gravitino mass,
the unity is related to the 3/2 gravitino components and the rest  to the 1/2 goldstino part. As an example in Fig.~\ref{fig:process_A} we present the relevant diagrams for the process A of table~\ref{table:1}.
For the calculation of the spin 3/2 part in  the  amplitudes,  following~\cite{Rychkov:2007uq}, we have employed the gravitino 
polarization sum
\beq
\Pi^{\tiny 3/2}_{\m \n}(P) =  \sum_{i = \pm 3/2} \Psi^{(i)}_\m \, \overline{\Psi}^{(i)}_\n = -\frac{1}{2} \g_\m \slash{P} \g_\n -  \slash{P} g_{\m\n}  \, ,
\label{eq:polsum}
\eeq
where $\Psi_\m$ is the gravitino spinor and $P$ its momentum. As in~\cite{Rychkov:2007uq}, for the goldstino spin 1/2 part   the 
nonderivative approach is used~\cite{Moroi:1995fs,Pradler:2007ne}. 
The result for the full squared amplitude  has been proved to be the same, either in the derivative or the nonderivative approach~\cite{Lee:1998aw}.

\begin{figure}[t]
\centering
\includegraphics[width=0.27\textwidth]{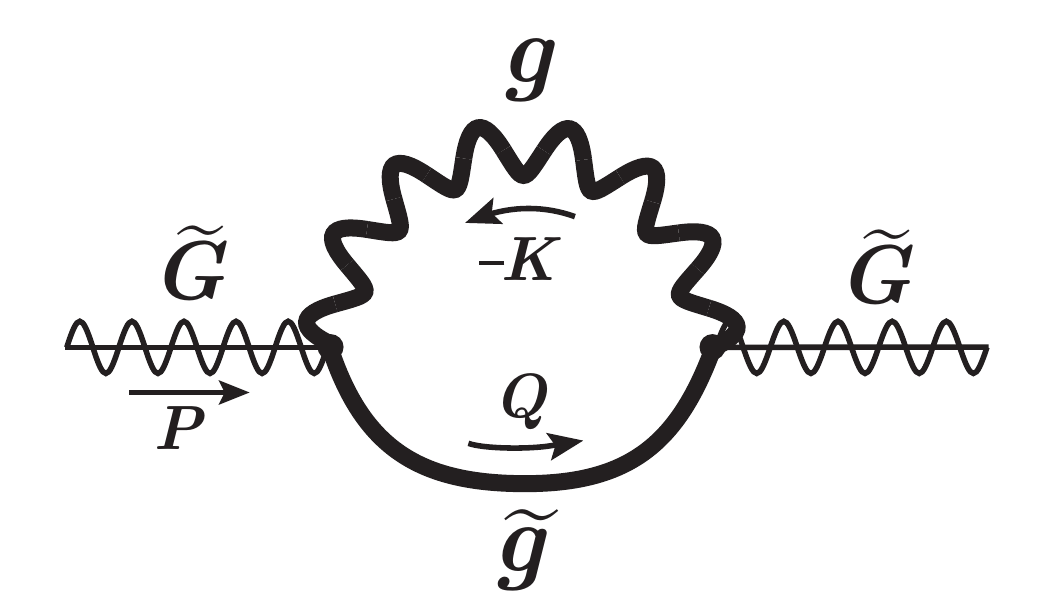}
\caption{The one-loop thermally corrected gravitino self-energy ($D-$graph) for the case of $SU(3)_c$. 
The thick gluon and gluino lines denote resummed thermal propagators. The  $SU(2)_L$ and $U(1)_Y$ cases have been also taken into account in our calculation.}
\label{D_GRAPH}
\end{figure}

The Casimir operators in Table~\ref{table:1} 
are $ C_N =\sum_{a,b,c} |f^{abc}|^2= N(N^2-1)=\lbrace 0,6,24\rbrace$ and
$C_N' =\sum^{\phi}_{ a,i,j} |T^a_{ij}|^2 = \lbrace   11,21,48 \rbrace$,
where $\sum^{\phi}_{ a,i,j}$ denotes the sum over all involved chiral multiplets and  group indices.
$f^{abc}$ and $T^a$ are the group structure constants and generators, respectively.  
Processes A, B and F are not present in  $U(1)_Y$ because $C_1=0$.
 The  masses for the particles  $a$, $b$ and $c$ are assumed to be zero. 
In the third  column of Table \ref{table:1} we present 
for each process the square of the full amplitude, which is the sum of individual amplitudes,
\beq
| {\cal M}_{X,\rm full} |^2 = |{\cal M}_{X,s} + {\cal M}_{X,t} + {\cal M}_{X,u} + {\cal M}_{X,x} |^2 \, ,
\label{Mfull}
\eeq
where the indices $s$, $t$, $u$ indicate the diagrams which are generated by the exchange of a particle in the corresponding channel and 
the index $x$ stands for the diagram  involving a quartic  vertex. The so-called 
$D-$graph, following the 
terminology of~\cite{Rychkov:2007uq},  is  illustrated in Fig.~\ref{D_GRAPH} for the case of the gluino-gluon  loop. 
Its contribution is the sum of the squared amplitudes for the $s$, $t$ and $u$ channel graphs,
\beq
|{\cal M}_{X,D}|^2 = |{\cal M}_{X,s} |^2  +  | {\cal M}_{X,t}|  ^ 2 +  | {\cal M}_{X,u} |^2 \, , 
\label{M2D}
\eeq
plus $1 \to 2$ processes.
This can be understood by  applying  the optical theorem. Hence,   from the imaginary part of the  loop graphs one  computes the sum of the decays ($1 \to 2$) and the scattering amplitudes ($2 \to 2$). In our case, we use resummed  thermal propagators for the gauge boson and gaugino and by applying cutting rules one sees that a $D-$graph describes both the scattering amplitudes appearing in~\eqref{M2D} and decay amplitudes.

The subtracted part of the squared amplitudes is the difference between the full amplitudes~\eqref{Mfull} and the amplitudes already included in the $D-$graph~\eqref{M2D}, i.e.  
\beq
|{\cal M}_{X,\rm sub}|^2 =  | {\cal M}_{X,\rm full} |^2 - | { \cal M}_{X,D} |^2 \, .
\label{M2sub}
\eeq
For the processes B,  F,  G, and H, the corresponding amplitudes  are  IR divergent. 
For this reason  we follow  the  more elegant method comprising the separation of 
the total scattering rate into two parts, the subtracted  and the $D-$graph part. It is worth to mention that
for the processes with incoming or/and outgoing gauge bosons, we have checked  explicitly the gauge invariance  for  $|{\cal M}_{X,\rm full}|^2 $.
On the other hand,  we  note that  $|{\cal M}_{X,\rm sub}|^2$ is gauge dependent. As it has been pointed out in~\cite{Rychkov:2007uq} the splitting 
of the amplitudes in resummed and non-resummed contributions violates the  gauge invariance. Therefore a gauge 
dependence of the result is  expected.

To sum up, the gravitino production rate $\gamma_{\tiny 3/2}$ consists of three  parts: (i) the subtracted rate $\gamma_{\rm sub}$ (ii) the  $D-$graph contribution  $\gamma_{\rm  D}$ 
and 
(iii) the top Yukawa rate $\gamma_{\rm top}$, 
\beq
\gamma_{\tiny 3/2} = \gamma_{\rm sub} + \gamma_{ \rm D} +\gamma_{\rm top}\,.
\label{gamma}
\eeq
Below,    these three contributions are  discussed  in detail.
\subsection{The subtracted rate}
In the fourth  column of Table \ref{table:1} we present the so-called subtracted part~(\ref{M2sub}), 
which is the sum of the interference terms among the four types of diagrams ($s$, $t$, $u$, $x$),   plus the $x$-diagram squared, for each process.
The subtracted part is nonzero only for the processes A and B.
Note that in~\cite{Rychkov:2007uq} the subtracted part for the processes H and J is also nonzero;  we assume that 
the authors had used  the squark-squark-gluino-goldstino   Feynman rule  as given  in~\cite{Bolz:2000xi}, where a factor $\gamma _5$
  is indeed  missing. In contrast, we are using the correct Feynman rule as given in~\cite{Pradler:2007ne}. 

To calculate the subtracted rate for the processes  
 $a\,b \rightarrow c\, \widetilde{G}$, we use the general form 
 \beq
\begin{aligned}
 \g = \frac{1}{(2\pi)^8} \int &  \frac{{\rm d^3} \mathbf{p}_a}{2E_a}  \, \frac{{\rm d^3} \mathbf{p}_b}{2E_b} \, \frac{{\rm d^3} \mathbf{p}_c}{2E_c} \, 
 \frac{{\rm d^3} \mathbf{p}_{\widetilde{G}}}{2E_{\widetilde{G}}} \, \,  |{\cal M}|^2\,  f_a \, f_b \, (1 \pm f_c)
 \, \,   \d^4(P_a + P_b - P_c  - P_{\widetilde{G}} )   \, ,
\label{collision_term1}
\end{aligned}
\eeq
where $f_i$ stands for the usual Bose and Fermi statistical densities 
\beq
f_{B|F} = {1 \over e^{E \over T} \mp 1}\, .
\eeq
In the temperature range of interest, all particles but the gravitino  are in thermal equilibrium. For the  gravitino the
statistical factor  $f_{\tiny  {\widetilde{G}}}$ is negligible. 
Thus, $1 -  f_{\tiny {\widetilde {G}}}  \simeq 1$,  as it  is already used in~\eqref{collision_term1}. 
 Furthermore,   backward reactions are neglected. 
In addition,    the simplification $1 \pm f_c  \simeq 1$  is usually applied,
making the analytic calculation of~\eqref{collision_term1}   possible.
In our case there is no such reason. We 
keep the factor  $1 \pm f_c$  and   consequently  
we proceed   calculating   the subtracted rate numerically. As it was argued in~\cite{Bolz:2000xi} and we have checked numerically the effect of taking into account the statistical factor $f_c$ can be around $-10$\%  ($+20$\%) if $c$ is a fermion (boson).

The  contribution of the  processes A and B, for each gauge group,   can be read  
from  Table~\ref{table:1} as
\beq
|{\cal M}_{A , \rm sub}|^2  + |{\cal M}_{B, \rm sub}|^2    =  \frac{g_N^2}{\mplanck ^2}\left(1+ \frac{m^2_{\lambda_N}}{3 m^2_{\tiny 3/2}} \right) C_N (-s + 2 t) \, .
\label{sub_A} 
\eeq
In~\eqref{sub_A}, a factor $1/2$ is already included for the process A due to the two identical incoming 
particles. Substituting ~(\ref{sub_A})   in ~(\ref{collision_term1}), the subtracted rate is obtained as  
\beq
\gamma _{\mathrm{sub}}=\frac{T^6}{\mplanck ^2} \sum_{N = 1}^3 g_N^2 \left(1+ \frac{m^2_{\lambda_N}}{3 m^2_{\tiny 3/2}} \right) C_N \left( -{\cal C}_{\scriptscriptstyle \rm BBF}^s +2\, {\cal C}_{\scriptscriptstyle \rm BFB}^t \right)\,.\label{sub:part}
\eeq
The numerical factors, calculated by using the Cuba library~\cite{Hahn:2004fe}, are $ {\cal C}_{\scriptscriptstyle \rm BBF}^s = 0.25957 \times 10^{-3} $ and 
${\cal C}_{\scriptscriptstyle \rm BFB}^t = -0.13286 \times 10^{-3}\,. $ The subscripts B and F specify if the particles are bosons or fermions, 
respectively, and the superscripts determine if the squared amplitude is proportional to $s$ or $t$. It is easy to see that our result 
for the subtracted part unlike in~\cite{Rychkov:2007uq} is negative. This   is not unphysical,  since the total rate and not the subtracted one is bound to be positive.
\subsection{The $D-$graph contribution}
 As it has been discussed above,   Eq.~\eqref{M2D} describes  
 the relation  between  the $D-$graph  and the sum of the squared amplitudes for  the $s$, $t$, and $u$ channels. 
In the $D-$graph contribution,  we will implement the resummed thermal corrections to the gauge boson  and gaugino  propagators. Like in~\cite{Rychkov:2007uq} using  the gravitino polarization sum~\eqref{eq:polsum}, we nullify   the corresponding quark-squark $D-$graph.
Although in Fig.~\ref{D_GRAPH}  the gluino-gluon thermal  loop is displayed, 
  the contributions of all the  gauge groups have been included in our analysis.
 The momentum flow used  to calculate the $D-$graph can be depicted in   Fig.~\ref{D_GRAPH}. That is
    $\Gr(P) \to g(K) + \sg(Q)$,   with 
$ P = (p, p, 0, 0)\,, \,  K = (k_0, k \cos\theta_k, k \sin\theta_k, 0)\,\, \text{and} \,\, Q = (q_0, q \cos\theta_q, q \sin\theta_q, 0)$,
where $\theta_{k,q}$ are the polar angles of the corresponding 3-momenta $\mathbf{k},\mathbf{q} $ in spherical coordinates.

\begin{figure}[t]
\centering
\includegraphics[width=0.483\textwidth]{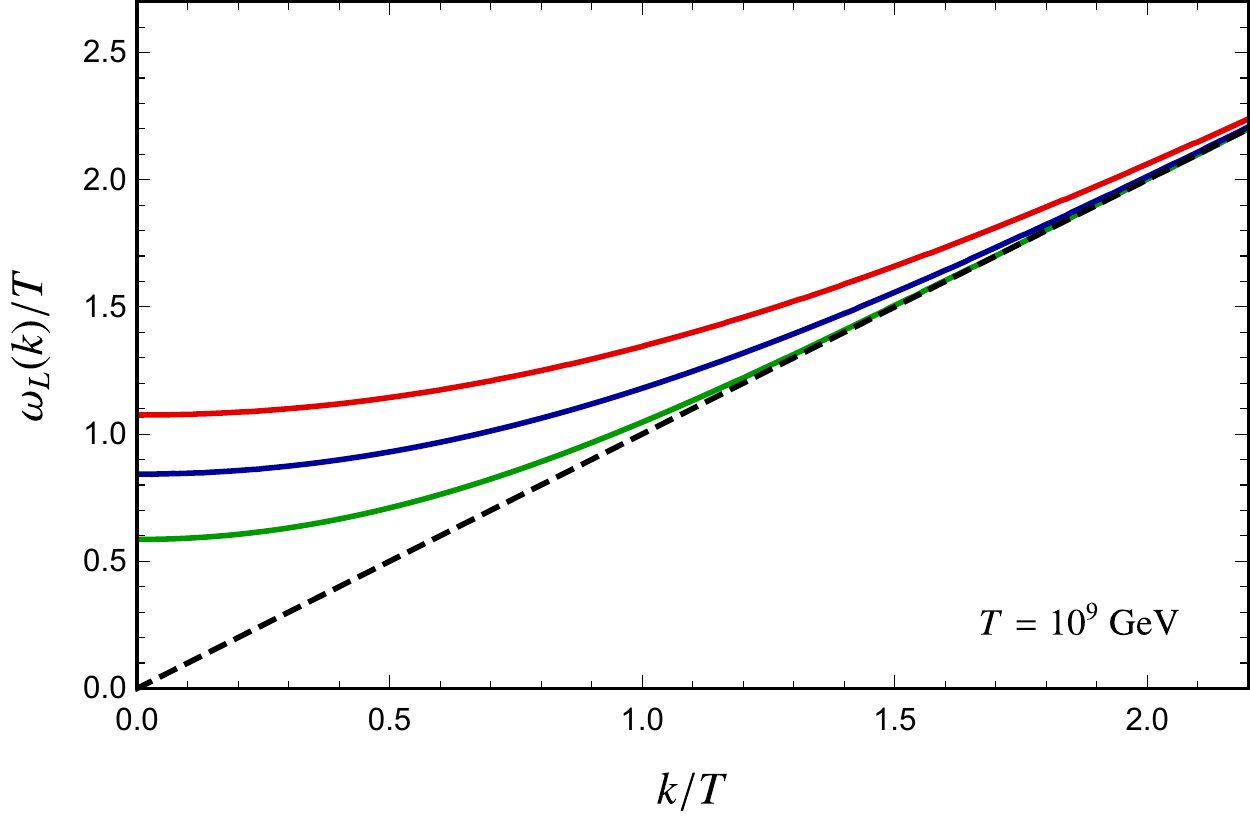}
\includegraphics[width=0.483\textwidth]{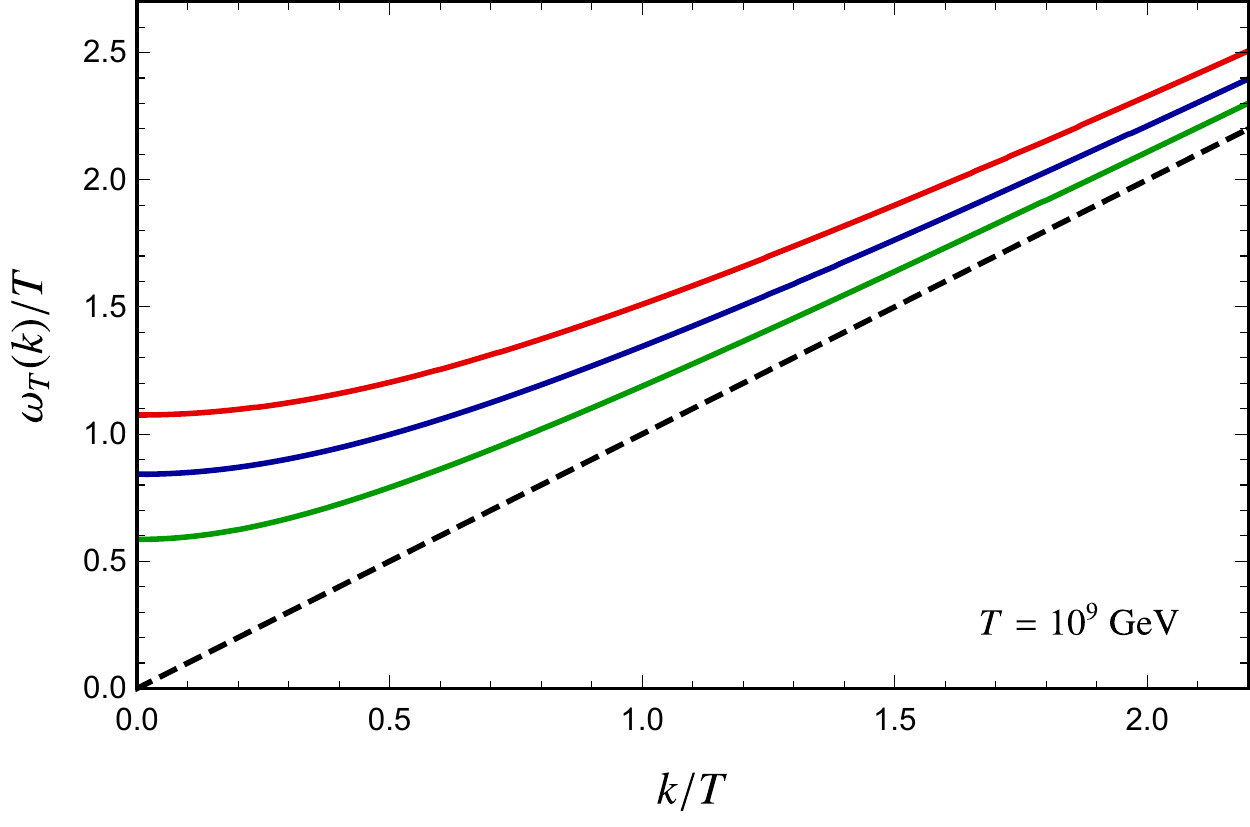}
\caption{Dispersion relations for longitudinal (left) and transverse (right) vector bosons. The colored curves represent the $SU(3)_c$ (red), $SU(2)_L$ (blue), and  $U(1)_Y$ (green) cases respectively. The dashed black line is the light-cone boundary. In both cases we have chosen $T=10^9\,\GeV$. } 
\label{fig:ome_LT}
\end{figure}

The  non-time-ordered  gravitino self-energy $\Pi^{<}(P)$ can be expressed in terms of the thermally resummed gaugino ${}^*S^<(Q)$ 
and gauge boson ${}^*D^<_{\m\n}(K)$ propagators as~\cite{Bolz:2000fu,Rychkov:2007uq}
\begin{eqnarray}
\Pi^<(P)&=& {1 \over 16 M^2_P} \sum_{N = 1}^3 n_N \left(1 + {m^2_{\lambda_N} \over 3 m^2_{\tiny 3/2}}\right) \int {{\rm d}^4 K \over (2 \pi)^4} \,
{\rm Tr}\left[ \slash{P} [\slash{K}, \g^\m] \, {}^*S^<(Q)  [\slash{K}, \g^\n] \, {}^*D^<_{\m\n}(K) \right]\, ,
\end{eqnarray}
where
\begin{eqnarray}
{}^*S^<(Q) &=&{ f_F(q_0) \over 2} \left[ (\g_0 - {\boldsymbol{\g}} \cdot \mathbf{q}/q) \, \r_+(Q) + (\g_0 + {\boldsymbol{\g}} \cdot \mathbf{q}/q) \, \r_-(Q)\right]\,, \\ \nn
{}^*D^<_{\m\n}(K) & = &  f_B(k_0) \left[ \Pi_{\m\n}^T \,  \r_T(K) + \Pi_{\m\n}^L  {k^2 \over K^2}\, \r_L(K) +\xi {K_\m K_\n \over K^4}\right]\, ,
\label{propagators}
\end{eqnarray}
with  $\xi$ being the gauge parameter, taken $\xi=1$
 in our
calculation and $n_N= \lbrace 1,3,8 \rbrace$. The spectral densities for longitudinal and transverse bosons are given by
\beq\label{eq:spectralL}
\rho_L(K)= -2{\rm Im} \frac{K^2}{k^2}\frac{1}{K^2-\pi_0 -\pi_L} \quad\quad\quad \text{and} \quad\quad\quad  \rho_T(K)= -2{\rm Im} \frac{1}{K^2-\pi_0 -\pi_T}\,,
\eeq
while the fermionic ones are
\beq\label{eq:rho_pm}
\r_\pm(Q)  =  -{\rm Im}\left\{\frac{q_0\mp q}{2} \left( 1 + \frac{1}{ 2 \pi^2} \Big[ {m_F^2 \over T^{2}}\Big(2  - \ln{-Q^2 \over \mu^2} \Big) \Big]\right) +  {m_F^2 \over T^{2}} F_\pm \right\} ^{-1}\,.
\eeq
\begin{figure}[t]
\centering
\includegraphics[width=0.483\textwidth]{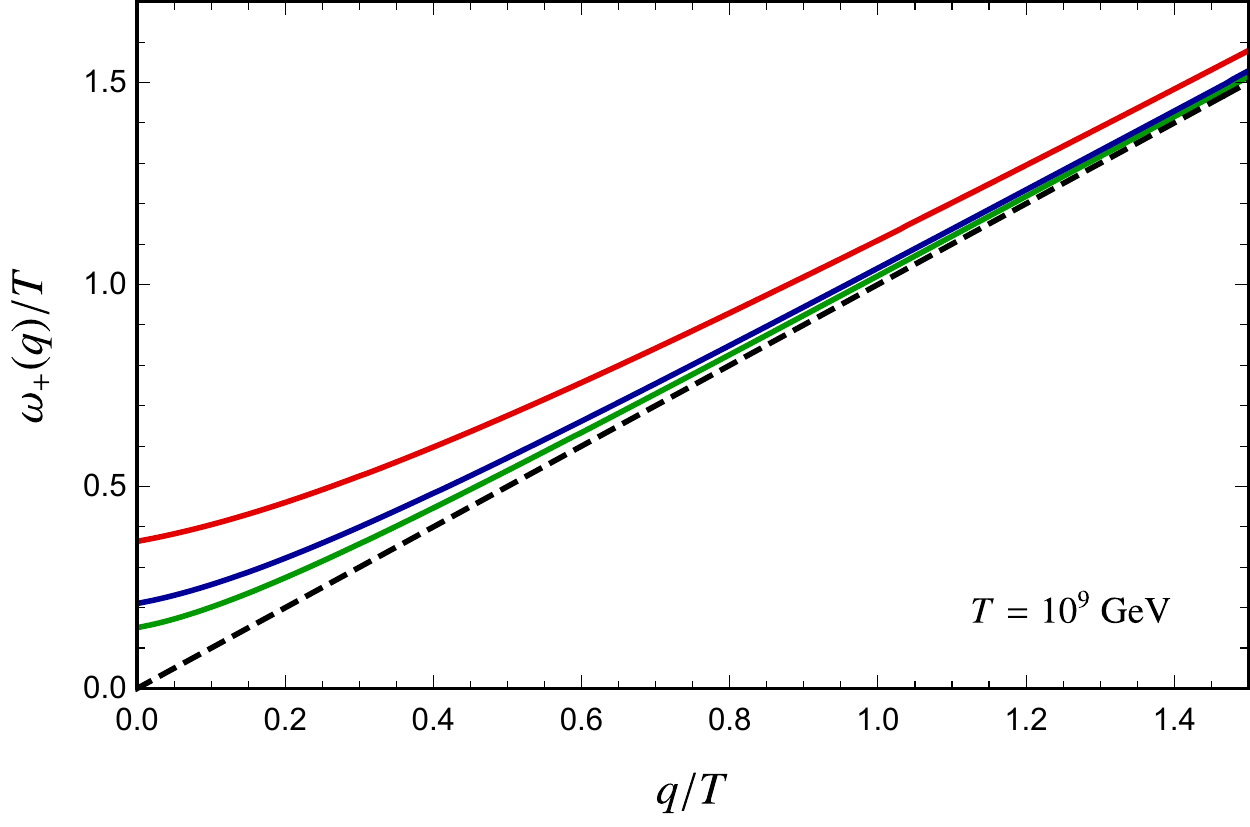}
\includegraphics[width=0.483\textwidth]{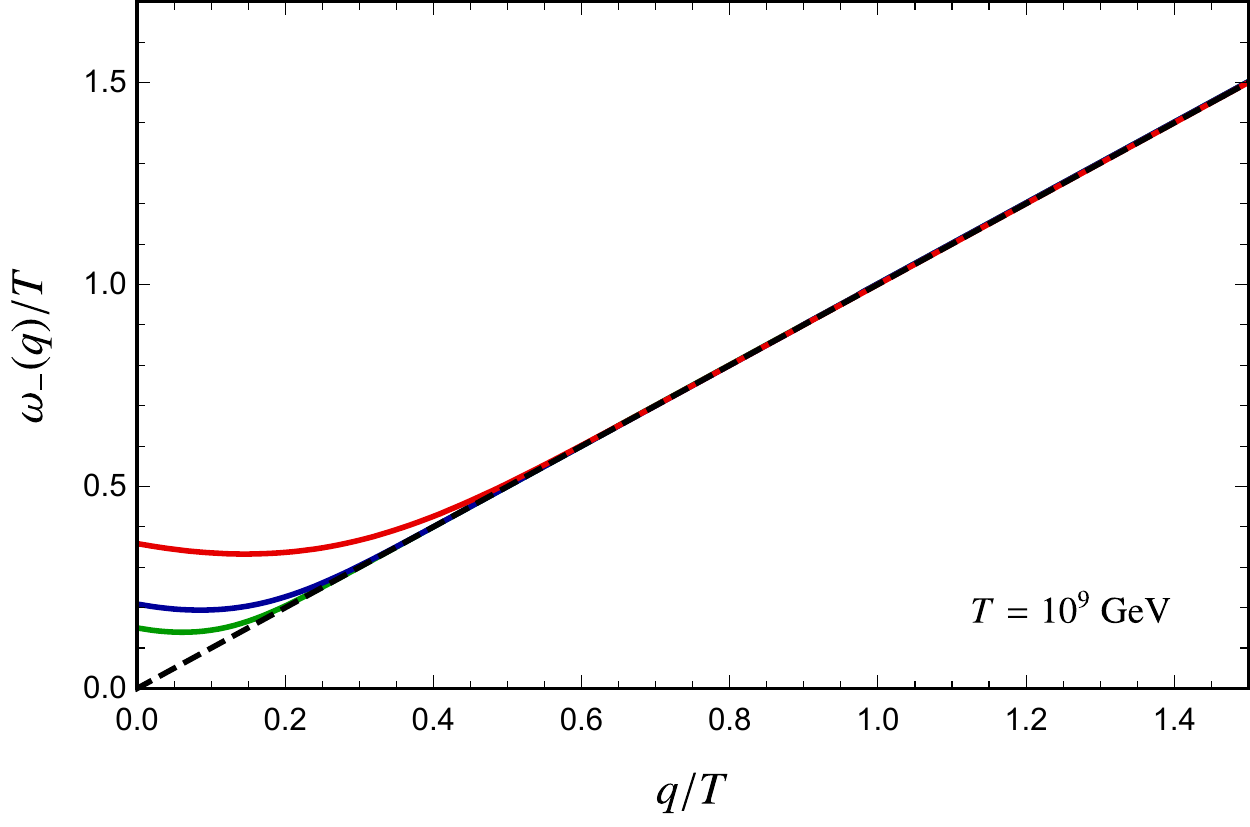}
\caption{Dispersion relations for fermions (left) and fermion holes (right). The colored curves represent the $SU(3)_c$ (red), $SU(2)_L$ (blue), and  $U(1)_Y$ (green) cases respectively. The dashed black line is the light-cone boundary. In both cases we have chosen $T=10^9\,\GeV$. } 
\label{fig:ome_pm}
\end{figure}
The  longitudinal and the transverse projectors $\Pi_{\m\n}^L$,  $\Pi_{\m\n}^T$, as well as the functions $\pi_L$, $\pi_T$, $\pi_0$, $F_\pm$ and the thermal mass $m_F$ can be found in Appendices B and C of~\cite{Rychkov:2007uq}. The spectral densities~(\ref{eq:spectralL}),(\ref{eq:rho_pm}) contain Landau damping contributions stemming from the imaginary part of the self-energies below and above\footnote{In the HTL approximation the Landau damping contributions are restricted only below the light-cone.} the light-cone and also  quasi-particles contributions for $k_0^2>k^2$ (and $q_0^2>q^2$). The quasi-particles contributions have the form  $\delta(k_0 \pm \omega_{L,T})$ for the bosons and $\delta(q_0 \pm \omega_{\pm})$ for the fermions, and are multiplied by suitable residues (see~\cite{Bellac:2011kqa} for the analytical expressions of the residues in the HTL approximation). The positions of the poles $\omega_{L,T}$ and $\omega_{\pm}$ can be very well approximated using the HTL limit, as it is discussed in~\cite{Rychkov:2007uq,Peshier:1998dy}. This argument is indeed true as we have seen numerically, but in order to be completely accurate, in Figs.~\ref{fig:ome_LT} and~\ref{fig:ome_pm} we display the dispersion relations for bosons and fermions, for the three gauge groups, beyond the HTL approximation.

 To compute the production rate related to the 
 $D-$graph  $ \gamma_D $,  we will use its definition~\cite{Bellac:2011kqa} 
\begin{equation}
\gamma_D = \int{{\rm d^3 \mathbf{p}} \over 2 p_0 (2 \pi)^3} \,\,  \Pi^<(p)\, 
\label{def_gamma_D_dp}
\end{equation}
and after appropriate  manipulations~\footnote{In principle, we use the $\d$-functions to reduce the number of the phase-space integrations. Further details will appear in~\cite{Eberl:20020gr}.}, we obtain
\begin{eqnarray}
\gamma_{\rm D} &=& {1 \over 4(2\pi)^5  \mplanck ^2} \sum_{N = 1}^3 n_N \left(1 + {m^2_{\lambda_N} \over 3 m^2_{\tiny 3/2}}\right)
 \int_0^\infty {\rm d} p \int_{-\infty}^\infty {\rm d } k_0  \int_0^\infty {\rm d} k \int_{|k-p|}^{k+p} {\rm d} q \,\, k \, f_B(k_0)  \, f_F(q_0)\label{gammadgraph} \\ \nn 
 &&  \times  \Big[ \r_L  (K) \, \r_- (Q)  \, (p - q)^2  \big( (p + q)^2 - k^2 \big) + \r_L (K) \, \r_+ (Q) \,  (p + q)^2  \big( k^2 - (p - q)^2  \big)\\ \nn  
           &&  + \, \,  \r_T  (K) \, \r_- (Q) \,   \big( k^2 - (p - q)^2  \big)   \Big( (1 + k_0^2 / k^2  \big)   \big( k^2 + (p + q)^2  \big) - 4 k_0 (p + q)  \Big) \\ \nn 
           && +  \, \, \r_T  (K) \, \r_+ (Q)  \, \big( (p + q)^2 - k^2  \big)  \Big( (1 +  k_0^2 / k^2 )  \big( k^2 + (p - q)^2  \big) - 4 k_0 (p - q)  \Big) \Big] \, ,
\end{eqnarray}
where $ q_0=p-k_0\,. $ 
The spectral functions $\rho_{L,T}$ and $\rho_{\pm}$ can be found in Eq.~(3.7)  in~\cite{Rychkov:2007uq}. The thermally corrected one-loop  self-energy for
gauge bosons, scalars and fermions that we have used in calculating these spectral functions can be found in~\cite{Weldon:1982aq, Weldon:1982bn,Weldon:1989ys,Weldon:1996kb, Peshier:1998dy,Weldon:1999th}. 
Comparing~\eqref{gammadgraph} with   the corresponding analytical result given in Eqs.  (4.6) and (4.7) in~\cite{Rychkov:2007uq}, 
one can notice that they differ on the overall factor and on the number of independent phase-space integrations.  
Our analytical result has been checked using various  frames for the momenta flow into the loop.
\subsection{The top Yukawa rate}
The production rate resulting from the top-quark Yukawa coupling $\lambda _t$ is given by~\cite{Rychkov:2007uq}
\beq
\gamma_{\mathrm{top}}=\frac{T^6}{\mplanck ^2} \, 72 \,\,  {\cal C}_{\scriptscriptstyle \rm BBF}^s \,  \lambda _t^2  \left(1+ \frac{A_t^2}{3m^2_{\tiny 3/2}} \right)\,,
\label{gamma:top}
\eeq
where $A_t$ is the trilinear stop supersymmetry breaking soft parameter
and ${\cal C}_{\scriptscriptstyle \rm BBF}^s = 0.25957 \times 10^{-3}$.  
 Since this contribution stems from the process  squark-squark $\rightarrow$ Higgsino-gravitino, 
  only the numerical factor ${\cal C}_{\rm BBF}^s$ is involved.
\begin{table}[h!]
\tbl{\label{table:2} The values of the constants $c_N$ and $k_N$ that parametrize our result~\eqref{gamma:paramtrization}. 
Each line  corresponds in turn to the particular gauge group, $U(1)_Y$,  $SU(2)_L$ or $SU(3)_c$.   }
{\begin{tabular}{ccc}
\hline \hline \\[-3mm]
\,  Gauge group \quad \quad  & \quad     $c_N$\quad  &\quad \quad \quad \quad$k_N$\quad  \\[1.2mm]
\hline \\[-3mm]
\quad $U(1)_Y$\quad \quad \quad \quad& \quad41.937 \quad&\quad\quad\quad \quad 0.824 \quad\\
\quad $SU(2)_L$\quad \quad \quad \quad&\quad 68.228 \quad&\quad \quad\quad \quad1.008\quad \\
\quad $SU(3)_c$\quad \quad\quad \quad & \quad21.067 \quad&\quad \quad \quad \quad6.878\quad\\
 \hline \hline 
\end{tabular}}
\end{table}
\section{Parameterizing    the result}
\setcounter{equation}{16}
Following~\cite{Ellis:2015jpg} we  parametrize the results~\eqref{sub:part} and~\eqref{gammadgraph} using the gauge couplings $g_1, g_2$ and $g_3\,.$ Thus 
\begin{equation}
\gamma_{\mathrm{sub} }+ \gamma_{\mathrm{D}} =  
 {3  \zeta(3) \over 16  \pi^3 } \, \frac{ T^6}{\mplanck ^2} 
          \sum_{N = 1}^3 c_N \, g_N^2  \left(1 + {m^2_{\lambda_N} \over 3 m^2_{\tiny 3/2}}\right) \ln \left( {k_N \over g_N}\right),
\label{gamma:paramtrization}
\end{equation}
\begin{figure}[t]
\centering
\includegraphics[width=0.483\textwidth]{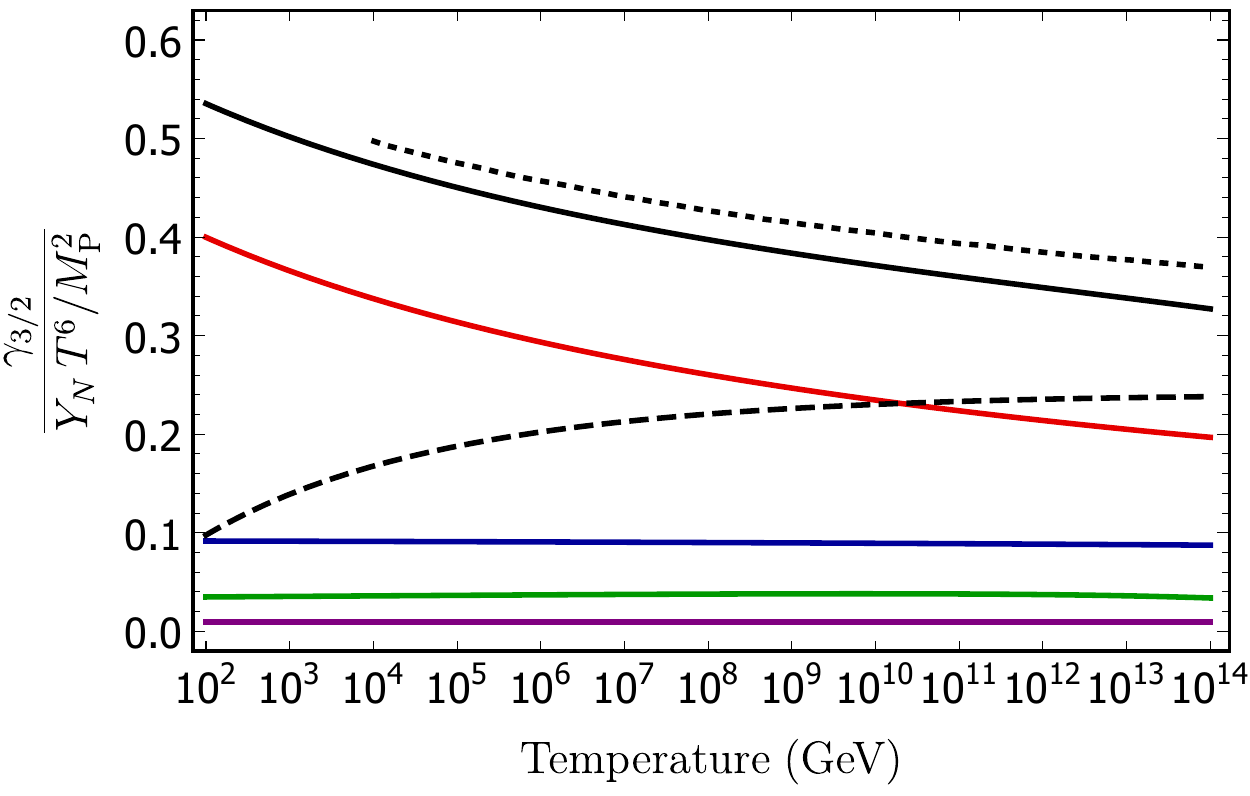}
\caption{The gravitino production rates in units of $Y_N\, T^6  / \mplanck ^2$. The solid curves represent in order
the total rate (black) given by~\eqref{gamma}, the $SU(3)_c$ (red), $SU(2)_L$ (blue), and  $U(1)_Y$ (green) rates given by~\eqref{gamma:paramtrization}, and the top Yukawa rate (purple) given by~\eqref{gamma:top}. The upper short dashed curve is the total production rate obtained in~\cite{Rychkov:2007uq}, while the lower
long dashed curve obtained in~\cite{Pradler:2006hh}. The top Yukawa coupling $\lambda _t$ has  been taken equal to 0.7 and the trilinear coupling $A_t$ has been ignored  so that our result can be directly compared with that in~\cite{Rychkov:2007uq}.} 
\label{fig:gamma}
\end{figure}
where the constants $c_N$ and $k_N$ depend on the gauge group and their values are given in Table~\ref{table:2}. 
In Fig.~\ref{fig:gamma} we summarize our numerical results for the gravitino production rates divided by $Y_N\, T^6  / \mplanck ^2$. 
Especially, for the case of the top Yukawa contribution,  in $Y_N$ the $m^2_{\lambda_N}$ has to be replaced by $ A_t^2 $.
The colored solid curves represent the $SU(3)_c$ (red), $SU(2)_L$ (blue), and  $U(1)_Y$ (green) rates given by~\eqref{gamma:paramtrization} 
and the top Yukawa rate (purple) given by~\eqref{gamma:top}, while the black solid curve is the total result given by~\eqref{gamma}. 
The dashed black curve  corresponds  to the  total result from~\cite{Rychkov:2007uq}. 
For an easy comparison, we have   chosen $\lambda _t = 0.7$. 
The  framework of our calculation is the  Minimal Supersymmetric Standard Model (MSSM) gauge group
structure $SU(3)_c\times SU(2)_L\times U(1)_Y$.
It is  interesting   that our result for the total gravitino production rate is only $5\%-11\% $ smaller  than  that in~\cite{Rychkov:2007uq}.  
We are not able to provide   a detailed  quantitative explanation.

For convenience, in Fig.~\ref{fig:gamma},  universal gauge coupling 
unification  is assumed at the Grand Unification (GUT) scale $\simeq  2\times 10^{16}\, \GeV$, 
but  the result in~\eqref{gamma:paramtrization} can be used in a more general framework.
Equation~\eqref{gamma:paramtrization} along with the numbers in Table~\ref{table:2} 
are  the main results of this talk. 
\section{Thermal Abundance of Gravitino}
\setcounter{equation}{17}
The relevant Boltzmann equation for the gravitino number density $n_{\tiny 3/2}$ can be written as 
\beq
\label{eq:boltzmann}
\frac{d{n}_{\tiny 3/2}}{dt} + 3H n_{\tiny 3/2} = \gamma_{\tiny 3/2}\,,
\eeq
where $H$ is the expansion rate of the Universe. The gravitino abundance $Y_{\tiny 3/2}$  is defined as the ratio of $n_{\tiny 3/2}$ to the number density of any single bosonic degree of freedom (DOF), $n_{\rm rad}=\zeta(3)T^3/\pi^2$. That is
\beq
Y_{\tiny 3/2}= {n_{\tiny 3/2} \over n_{\rm rad} }\,,
\label{eq:abundance}
\eeq
From~\eqref{eq:boltzmann} and~\eqref{eq:abundance} 
 we obtain that the gravitino {}abundance for temperatures much lower than the reheating temperature, i.e. $T \ll T_{\rm reh}$ is given by~\cite{Ellis:2015jpg}
\beq
Y_{\tiny 3/2}(T)\simeq {\gamma _{\tiny 3/2}(T_{\rm reh}) \over H(T_{\rm reh}) \,\,  n_{\rm rad}(T_{\rm reh}) }\,\, {g_{*s}(T) \over g_{*s}(T_{\rm reh})}\,,
\label{eq:abundance_app}
\eeq
where  $g_{*s}$ are the entropy density DOFs in the MSSM.
Notice that Eq.~\eqref{eq:abundance_app} is consistent with the assumption that the inflaton decays are instantaneous, as the Universe is thermalizing. The case of not instantaneous inflaton decay has been also studied in~\cite{Ellis:2015jpg,Garcia:2017tuj}\footnote{In particular, following~\cite{Garcia:2017tuj},  a correction   factor  $\sim 10\%$ is expected for not  instantaneous inflaton decays, in the case of gravitino DM.}.
\begin{figure}[t]
\centering
\includegraphics[width=0.483\textwidth]{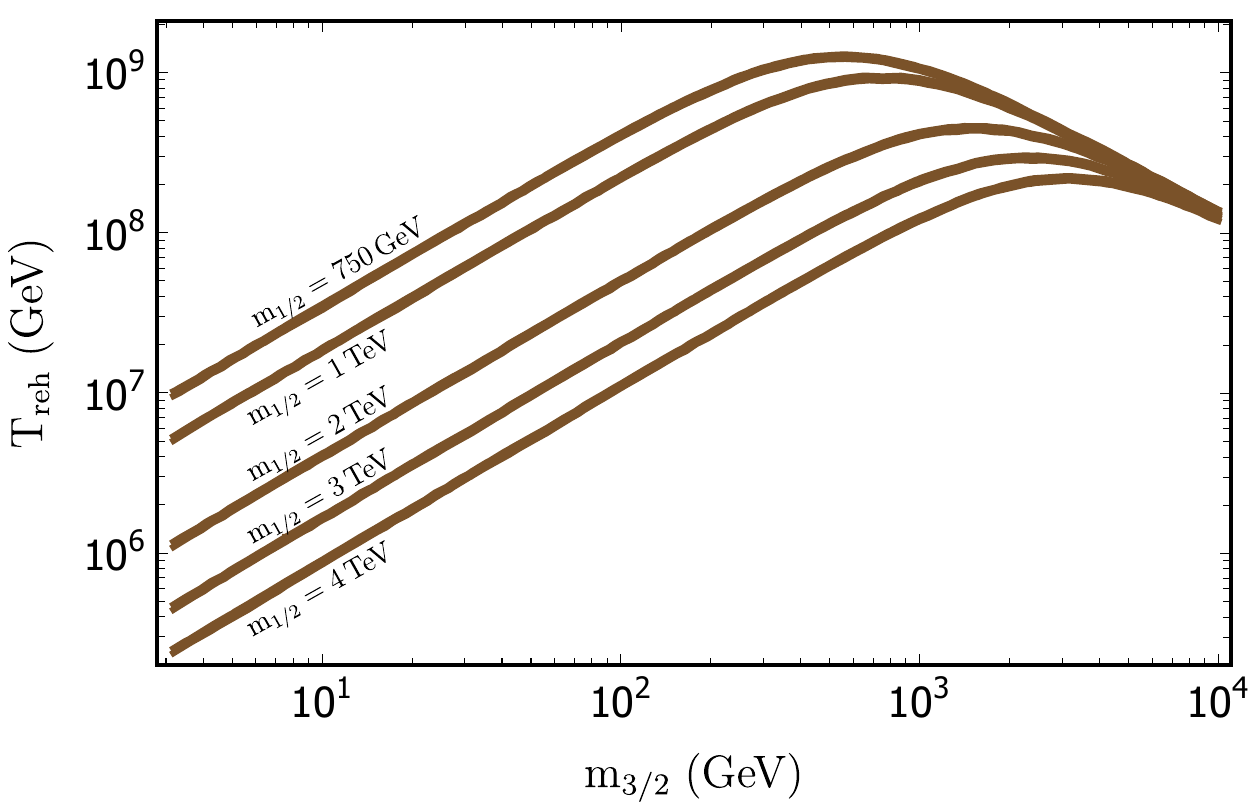}
\caption{The thick show the $3 \sigma$ regions where $\Omega _\mathrm{DM}$ is equal to the gravitino abundance for various values of $m_{\tiny 1/2}$. As previously  the top Yukawa coupling is $\lambda _t =0.7$ and the trilinear coupling $A_t$ has been ignored.}
\label{fig:Treh_m}
\end{figure}

According to the latest data {}from the Planck satellite~\cite{Aghanim:2018eyx}, the cosmological accepted value for the 
DM density in the Universe  is $\Omega _\mathrm{DM} h^2= 0.1198 \pm 0.0012$. 
Assuming that the entire thermal gravitino abundance is the  observed DM,  we obtain that
\begin{equation}
\Omega _\mathrm{DM} h^2 = {\rho_{\tiny 3/2}(t_0)  \,  h^2 \over \rho _{\rm cr} }
 \simeq  1.33\times 10^{24} \,  {m_{\tiny 3/2} \,\, \gamma_{\tiny 3/2}(T_{\rm reh}) \over T_{\rm reh}^5 }\,,
\label{eq:dmdensity}
\end{equation}
where $\rho_{\tiny 3/2}$ is the gravitino mass density, $\rho _{\rm cr}=3\, H_0^2 \mplanck ^2$ is the critical energy density,  $H_0=100\, h\, {\rm km/(s\, Mpc)}$ is
 the Hubble constant today,  and $T_0 =2.725\,K$   the temperature of the
 cosmic microwave background in the current epoch. Assuming the MSSM content, the entropy DOFs at 
the relevant temperatures are $g_{*s}(T_0)=43/11$ and $g_{*s}(T_{\rm reh})=915/4$. 
Figure~\ref{fig:Treh_m} illustrates the $3\sigma$  regions resulting from~\eqref{eq:dmdensity}
 for various
values of $m_{\tiny 1/2}$ between   $750\GeV$ and  $4\TeV$. In this, the top Yukawa coupling has been choosen to have the value $\lambda _t =0.7$ as previously, and the trilinear coupling $A_t$ has been ignored.
As before, a  gauge coupling  unification and a universal gaugino mass $m_{\tiny 1/2}$ are assumed at the GUT scale $\simeq  2\times 10^{16}\, \GeV$.

In the large gravitino mass limit, $m_{\tiny 1/2} \ll m_{\tiny 3/2}$, the 
characteristic factor $m_{\lambda_N}^2 / (3 m^2_{\tiny 3/2})$ becomes negligible and so the reheating temperature is $m_{\tiny 1/2}$ independent as it is shown in Fig.~\ref{fig:Treh_m}.
The recent LHC data on  gluino searches~\cite{Aaboud:2017hrg,Sirunyan:2019ctn}, suggest that $m_{\tiny 1/2} \gtrsim\textit{} 750 \, \GeV$, and thus according to Fig.~\ref{fig:Treh_m} the maximum reheating temperature $T_\mathrm{reh}\simeq  10^9\, \GeV$ corresponds to a gravitino mass $m_{\tiny 3/2} \simeq 550\, \GeV $. 
For the same gravitino mass, $m_{\tiny 1/2}$ can go up to $3-4 \TeV$,  for a reheating temperature an order of magnitude smaller  $T_\mathrm{reh} \simeq  10^8 \GeV$.

\section{Conclusions}
In this talk, we have presented a calculation of  the gravitino thermal abundance, using 
the full one-loop thermally corrected  gravitino self-energy~\cite{Eberl:2020fml}. 
Having amended   the main analytical 
formulas  for the gravitino production  rate,  we  have computed it numerically without any approximation.
In addition we provide  a simple and useful parametrization of our final result.
Moreover, in the context of minimal supergravity models, 
assuming gaugino mass unification, we have updated the bounds on  the
 reheating temperature for certain gravitino masses. In particular,  saturating the current  LHC  gluino mass
  limit $m_{\tilde g} \gtrsim 2100 \GeV$, we find that  a  maximum reheating temperature 
$T_\mathrm{reh} \simeq  10^9 \GeV$  is compatible to  a gravitino mass $m_{\tiny 3/2}  \simeq 500 - 600\, \GeV $. 

As for the  constraints on  the reheating temperature, we have  applied   
 the cosmological  data on   gravitino DM scenarios. This   
tells  us whether  thermal leptogenesis is a possible  mechanism for generating baryon asymmetry 
of the Universe. Successful  thermal leptogenesis requires high temperature,
 $T_\mathrm{reh} \gtrsim 2\times  10^9\, \GeV$~\cite{Giudice:2003jh,Antusch:2006gy,Buchmuller:2005eh}, which is marginally bigger than the maximum  reheating temperature obtained in our model using the lowest $m_{\tiny 1/2}  $ mass demonstrated in the  recent  LHC  data~\cite{Aaboud:2017hrg,Sirunyan:2019ctn}.  
In any case, there are many alternative models  for baryogenesis. 
In addition, as it has been pointed out before, the thermal gravitino abundance is in general   a part of the 
whole DM density and the inclusion of  other  components will affect  the  phenomenological analysis.

\section*{acknowledgments}
This research is co-financed by Greece and the European Union (European Social Fund- ESF) through the  Operational Programme \textquote{Human Resources Development, Education and Lifelong Learning} in the context of the project \textquote{Strengthening Human Resources Research Potential
via Doctorate Research - 2nd Cycle} (MIS-5000432), implemented by the State Scholarships Foundation (IKY).  This research work was supported by the Hellenic Foundation for Research 
and Innovation (H.F.R.I.) under the ``First Call for H.F.R.I. Research Projects to support 
Faculty members and Researchers and the procurement of high-cost research equipment grant'' (Project Number: 824).


\end{document}